\documentclass[twocolumn,showpacs,preprintnumbers,prl]{revtex4}
\usepackage{amsfonts}
\usepackage{amsmath}
\usepackage{amssymb}
\usepackage{graphicx}
\usepackage{hyperref}
\usepackage[title,titletoc,header]{appendix}

\begin{document}

\title{Induced transparency in optomechanically coupled resonators}

\author{Zhenglu Duan$^{1}$, Bixuan Fan$^{1}$, Thomas M. Stace$^{2}$, G. J. Milburn$^{2}$ and Catherine A. Holmes$^{3\ast}$}
\affiliation{$^{1}$College of Physics, Communication and Electronics,
Jiangxi Normal University, Nanchang, 330022, China}
\affiliation{$^{2}$Center for Engineered Quantum Systems, School of Mathematics and Physics,
The University of Queensland, St Lucia, Queensland 4072, Australia}
\affiliation{$^{3}$School of Mathematics and Physics, University of Queensland, St Lucia, 4072,
Qld, Australia}
\keywords{}

\pacs{PACS number}
\pacs{42.65.Pc, 42.50.Wk, 42.50.Nn, 07.07.Df}

\begin{abstract}
In this work we theoretically investigate a hybrid system of
two optomechanically coupled resonators, which exhibits induced transparency. This is
realized by coupling an optical ring resonator to a toroid. In the
semiclassical analyses, the system displays bistabilities, isolated branches
(isolas) and self-sustained oscillation dynamics. Furthermore, we find that
the induced transparency transparency window sensitively relies
on the mechanical motion. Based on this fact, we show that the described
system can be used as a weak force detector and the optimal sensitivity can beat the standard quantum limit without using feedback control or squeezing under available experimental conditions.
\end{abstract}

\maketitle
\footnotetext[1]{$^\ast$Corresponding author, cah@maths.uq.edu.au}

\textit{Introduction.} The coupled-resonator induced transparency (CRIT),
arising when light passes through two closely located identical ring optical
resonators, was first put forward in \cite{PhysRevA.69.063804} and then
experimentally observed in \cite{PhysRevA.71.043804,PhysRevLett.96.123901}.
This transparency comes from the interference between two pathways of the
light, leading to the Fano effect in the coupled ring resonator system. The CRIT effect
has a wide variety of applications in linear regime, such as slowing and
stopping light \cite{PhysRevLett.98.213904,Xu:06}, signal routing \cite%
{Mancinelli:11}, and biomedical molecule sensing \cite{Liu2009Planar}. In the highly nonlinear regime, the system shows bistability \cite%
{Lu2008Proposal}, which can be used for optical switching \cite{Maes:05}.

Here we consider such a situation: One of the ring resonators in CRIT is
replaced by an optomechanical system (toroidal), in which the optical mode and mechanical vibration
mode are coupled through radiation pressure coupling. This simple structure
forms a hybrid CRIT and optomechanical system. As we show the interplay
between CRIT and optomechanics can lead to more interesting effects. In
particular by including the optomechanical nonlinearity the transparency
window of the CRIT becomes tunable and also exhibits bistability. In
addition the coupled system undergoes saddle node and Hopf bifurcations
which result in the emergence of bistabilities, isolated branches (isolas)
\cite{ganapathisubramanian1984bistability,chen2010classical} and
self-sustained oscillations.

Detection of weak forces with high sensitivity has long been a research
focus. Optomechanical systems are ideal for detecting small perturbations
including weak forces \cite%
{PhysRevA.64.051401,PhysRevA.74.063816,Teufel2009Nanomechanical,Gavartin2012A}%
. The general idea for realizing those proposals is to map a weak force onto
the shift in position of a mechanical system that can be easily monitored by
the coupled optical field. Also, precise quantum
measurements can be performed via monitoring transparency windows in EIT or
its analog, the optomechanically induced transparency (OMIT) \cite%
{Weis2010Optomechanically,Safavi2011Electromagnetically,Fan2015Cascaded}, for instance,
measurements of magnetic fields \cite{PhysRevLett.69.1360,Lee1998Sensitive},
Rydberg states \cite{Mohapatra2007Coherent}, number of electric charges \cite%
{Zhang2012Precision} and transition dipole moments \cite{Qi2002Measurement}.
\begin{figure}[tb]
\centering
\includegraphics[width=2.5in]{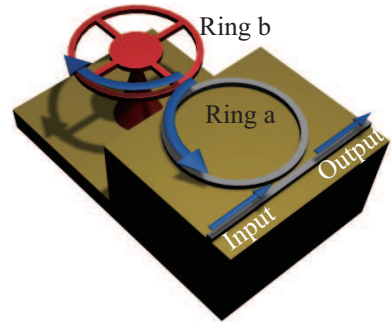}
\caption{{\protect\footnotesize (Color online) Schematic for the
coupled-resonator system with optomechanical coupling. An optical ring
resonator (the green ring) is coupled to a toroid (the red ring)
evanescently. In the toroid, the optical mode and mechanical mode interact
through radiation pressure.}}
\label{figure1}
\end{figure}
Stimulated by these possible applications we utilize the
large dispersion of the CRIT transparency window and the sensitive
dependence of CRIT on mechanical motion to detect a weak force applied on
the mechanical mode. With shot noise and environmental thermal noise included, we estimate that the optimal force sensitivity is $17$aN$%
\cdot $Hz$^{-1/2}$ when the system operates at resonance.

\textit{Model.} The system under consideration is shown in figure \ref%
{figure1}: a ring optical resonator is evanescently coupled to a toroidal optomechanical system and
a driving field is applied on the optical resonator through a waveguide. In
the frame rotating at the driving field frequency $\omega _{\text{in}}$, we
write down the equations of motion for the system operators: \label{tabpc}
\begin{eqnarray}
\dot{\hat{x}} &=&\omega _{m}\hat{p}  \label{x} \\
\dot{\hat{p}} &=&-\gamma _{m}\hat{p}-\omega _{m}{\hat{x}}+g_{1}\hat{b}^{\dag
}\hat{b}+\hat{\xi}_{m}  \label{p} \\
\dot{\hat{a}} &=&-(i\Delta +\kappa )\hat{a}-ig_{2}{\hat{b}}+\hat{a}_{\text{in%
}}  \label{a} \\
\dot{\hat{b}} &=&-(i(\Delta +\delta -g_{1}\hat{x})+\kappa _{b}/2){\hat{b}}%
-ig_{2}\hat{a}+\sqrt{\kappa _{b}}{\hat{b}}_{\text{vac}}  \label{b}
\end{eqnarray}%
where $\hat{x}$ and $\hat{p}$ are the dimensionless position and momentum
operators for the mechanical degree of freedom with frequency $\omega _{m}$.
$\hat{a}$ and $\hat{b}$ are annihilation operators of the optical modes in
the resonator and toroid, with frequencies $\omega _{a}$ and $\omega _{b}$
and damping rates $\kappa _{a}$ and $\kappa _{b}$. $\kappa =\left( \kappa
_{a}+\kappa _{\text{ex}}\right) /2$. $\Delta =\omega _{a}-\omega _{\text{in}%
} $ is the detuning between the driving light and the optical resonator $%
\hat{a}$ and $\delta =\omega _{b}-\omega _{a}$ is the frequency difference
between two optical modes. $g_{1}$ is the optomechanical coupling
coefficient between the optical mode $\hat{b}$ and the mechanical mode of
the toroid and $g_{2}$ is the coupling coefficient between two optical
modes. $\kappa _{\text{ex}}$ is the outgoing coupling coefficient from the
optical resonator into the waveguide. $\hat{a}_{\text{in}}=\sqrt{\kappa _{%
\text{ex}}}(\sqrt{I_{\text{in}}}+\delta \hat{a}_{\text{in}})+\sqrt{\kappa
_{a}}\hat{a}_{\text{vac}}$ where the driving light intensity $I_{\text{in}}$
with fluctuation $\delta {\hat{a}}_{\text{in}}$. $\hat{a}_{\text{vac}}$ and
${\hat{b}}_{\text{vac}}$\ are the external vacuum fields to optical modes a
and b, respectively. The mechanical mode is affected by a viscous force with
the damping rate $\gamma _{m}$ and by a Brownian stochastic force with noise
$\hat{\xi}_{m}$.

When the system is strongly driven, it can be characterized by the
semiclassical steady-state solutions with large amplitudes for both
mechanical and optical modes. In the following, we denote $y_{s}$ as the
steady-state mean value of the operator $\hat{y}$. By setting the time
derivatives of system variables to zero and factorizing the expectation
values, the steady-state solutions are obtained:
\begin{eqnarray}
x_{s} &=&\frac{g_{1}g_{2}^{2}\left\vert a_{s}\right\vert ^{2}}{\omega
_{m}(\Delta _{\text{eff}}^{2}+\kappa _{b}^{2}/4)}  \label{xs} \\
a_{s} &=&\frac{\sqrt{\kappa _{\text{ex}}I_{\text{in}}}}{(i\Delta +\kappa
+g_{2}^{2}/(i\Delta _{\text{eff}}+\kappa _{b}/2))}  \label{as} \\
b_{s} &=&-\frac{ig_{2}}{(i\Delta _{\text{eff}}+\kappa _{b}/2)}a_{s}
\label{bs}
\end{eqnarray}%
with the effective detuning $\Delta _{\text{eff}}=\Delta +\delta -g_{1}x_{s}$%
. Eq. (\ref{xs}) is a cubic equation for the steady state value $x_{s}$,
therefore there are at most three real roots.

Next, we turn to study fluctuations around the steady state by expanding the
system operators around their stable-state values, i.e., $\hat{y}\rightarrow
y_{s}+\hat{y}$, and introducing the field quadrature fluctuations $\hat{X}%
_{c}=\left( \hat{c}+\hat{c}^{\dag }\right) /\sqrt{2}$ and $\hat{Y}%
_{c}=\left( \hat{c}-\hat{c}^{\dag }\right) /\left( \sqrt{2}i\right) $ ($\hat{%
c}=\hat{a},\hat{b},\hat{a}_{\text{in}},\hat{a}_{\text{vac}},\hat{b}_{\text{%
vac}}$). Ignoring high-order terms of fluctuations, the linearized equations
of motion can be written as $\dot{y}=J\hat{y}+\xi $ where $\hat{y}=\left[
\hat{x},\hat{p},\hat{X}_{a},\hat{Y}_{a},\hat{X}_{b},\hat{Y}_{b}\right] ^{T}$%
, the noise term $\xi =\left[ 0,\xi _{m},\sqrt{\kappa _{\text{ex}}}\hat{X}_{%
\text{in}}+\sqrt{\kappa _{a}}\hat{X}_{\text{vac}}^{a},\sqrt{\kappa _{\text{ex%
}}}\hat{Y}_{\text{in}}+\sqrt{\kappa _{a}}\hat{Y}_{\text{vac}}^{a},\right.\\
\left. \sqrt{\kappa _{b}}\hat{X}_{\text{vac}}^{b},\sqrt{\kappa _{b}}\hat{Y}_{%
\text{vac}}^{b}\right] ^{T}$, and the Jacobian matrix is given by
\begin{equation}
J=\left(
\begin{array}{cccccc}
0 & \omega _{m} & 0 & 0 & 0 & 0 \\
-\omega _{m} & -\gamma _{m} & 0 & 0 & g_{1}X_{bs} & g_{1}Y_{bs} \\
0 & 0 & -\kappa  & \Delta  & 0 & g_{2} \\
0 & 0 & -\Delta  & -\kappa  & -g_{2} & 0 \\
-g_{1}Y_{bs} & 0 & 0 & g_{2} & -\kappa _{b}/2 & \Delta _{\text{eff}} \\
g_{1}X_{bs} & 0 & -g_{2} & 0 & -\Delta _{\text{eff}} & -\kappa _{b}/2%
\end{array}%
\right) ,
\end{equation}%
where $X_{bs}=\left( b_{s}+b_{s}^{\ast }\right) /\sqrt{2}$, and $%
Y_{bs}=\left( b_{s}-b_{s}^{\ast }\right) /\left( \sqrt{2}i\right) $. The
steady-state solution is stable if all eigenvalues of the Jacobian matrix
have negative real parts. The Routh-Hurwitz criterion \cite%
{Gradshteyn1994Table} can be used to determine the stable and unstable
regions in the parameter space.

\begin{figure}[tb]
\includegraphics[width=3.7in]{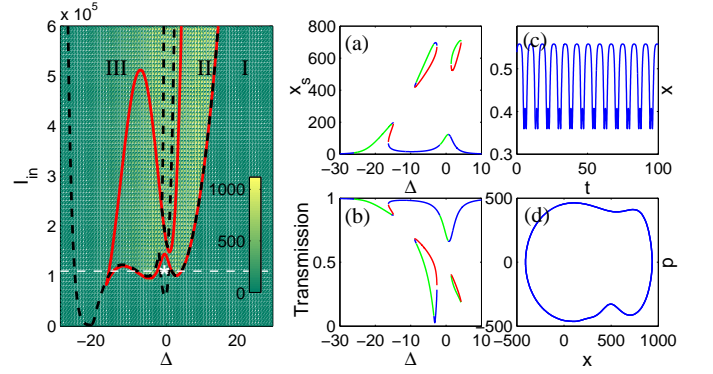}
\caption{{\protect\footnotesize (Color online) Left panel: The bifurcation
diagram for the position of the mechanical resonator in the ($\Delta ,I_{in}$%
) plane. The red solid curve labels the boundary of saddle-node bifurcations
and the black dashed curve labels the boundary of Hopf bifurcations. The
stable region is denoted by region I, which is the area outside of red and
black curves. The area within the red curve is region II, in which multiple
steady states exist. Region III, enclosed with black curve, is the
parametric unstable region, with self sustained oscillations created in the
Hopf bifurcation. In right panel, figure (a)is a cross section of left panel
labeled by the white dashed line. The blue, red and green curves denote
stable, unstable and parametric unstable solutions. Figure 2 (b) shows the
corresponding light transmission of (a). The parameters are $k_{a}=0.6$, $%
k_{b}=4.0$, $k_{\text{ex}}=4 $, $\protect\delta =20$, $g_1=0.03$, $g_2=4$, $%
\protect\omega _{m}=1 $ and $\protect\gamma_m = 0.1$. Figures 2 (c) and (d)
present the dynamics and its corresponding phase space picture for one point
in region III $(\Delta,I_{in})=(-0.25,1.1\times 10^{5})$, as labelled using
a white star in the left panel.}}
\label{figure2}
\end{figure}

To illustrate the stability of the described system, the phase diagram for
the mechanical position of the toroid in the parameter space ($\Delta ,I_{%
\text{in}}$) is pictured in left panel of figure \ref{figure2}. The whole
plane is divided into stable, unstable and parametric unstable regions,
which are divided by saddle-node bifurcations (red solid curve) and Hopf
bifurcations (black dashed curve). In particular, we plot a cross section of
the left figure in figure \ref{figure2} (a) as labeled by the white dashed
line ($I_{\text{in}}=1.1\times 10^{5}$). There is the typical bistable
branch on the main continuous curve and two isolated branches (isolas) above
the lower main branch. Supposing that initially the equilibrium position of the
mechanical resonator is on an isolated branch, if the detuning $\Delta$ is
continuously swept in an increasing or decreasing manner, the equilibrium
position will jump to the lower main branch, but not vice versa. This unique
feature is a non-hysteretic bistability, which can be used as a
unidirectional switch. In fact, the stability of the isolas are similar to
conventional S-type bistabilities in the higher-dimensional parameter space.
In experiments the light transmission $T\equiv \left\vert a_{\text{out}}/%
\sqrt{I_{\text{in}}}\right\vert ^{2}$ is more directly observable. According
to the input-output relation $a_{\text{out}}=\sqrt{I_{\text{in}}}-\sqrt{\kappa _{%
\text{ex}}}a_{s}$, the light transmission can be related to the mechanical
position as
\begin{equation}
T=\left\vert 1-\sqrt{\kappa _{\text{ex}}/I_{\text{in}}}a_{s}\right\vert ^{2}
\label{T}
\end{equation}
Figure 2 (b) presents the equilibrium curve of the light transmission
corresponding to figure 2(a) and one can observe that it also shows
hysteretic behaviors and the emergence of isolas.

The presence of the parametric unstable region comes from the choosing
parameters lying in unresolved sideband regime with small mechanical
damping, i.e., $1>\omega _{m}/\kappa _{b}>\gamma _{m}$. Figure 2(c) presents
the same in the parametric unstable region (III) where the mechanics
exhibits stable oscillatory motion. The corresponding phase space picture is
displayed in figure 2(d) and one can easily find a stable closed phase
trajectory ( a stable limit cycle).

\textit{CRIT-like effect.} If the optomechanical coupling is switched off,
the model reduces to a typical optical coupled-resonator system, which
manifests the CRIT effect for certain parameters \cite%
{PhysRevA.69.063804,PhysRevA.51.4959}. In the following we investigate how
the nonlinearity induced by the mechanical resonator changes the CRIT
effect. For simplicity, we consider two optical resonators with the same
frequency, i.e., $\delta =0$, the vanishing decay of the optical field in
the toroid, low driving intensity and $\kappa _{a}=\kappa _{\text{ex}}$.
Under these assumptions, Eq. (\ref{T}) becomes
\begin{equation}
T=\left[ 1+\kappa ^{2}/\left( \Delta -g_{2}^{2}/\left( \Delta
-g_{1}x_{s}\right) \right) ^{2}\right] ^{-1}  \label{CRIT}
\end{equation}%
which is a CRIT-like function with the transparency window width $g_{2}$ and
the center position of the transparency window located at $g_{1}x_{s}$.
Recall that $x_{s}$ is dependent on the detuning $\Delta $ and the input
light intensity $I_{\text{in}}$. Eq. (\ref{CRIT}) shows us how the presence
of the mechanical motion modifies the behaviors of CRIT.

\begin{figure}[tb]
\includegraphics[width=3.7in]{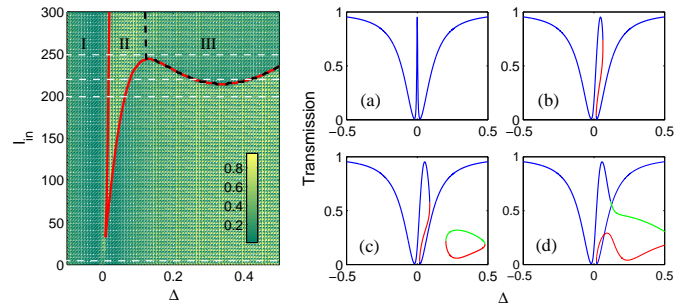}
\caption{{\protect\footnotesize (Color online) Left panel: the bifurcation
diagram for the position of the mechanical resonator in the ($\Delta ,I_{in}$%
) plane. Right panel: the transmission as a function of detuning $\Delta $
with different input light intensity. (a)$I_{\text{in}}=1$, (b)$I_{\text{in}%
}=200$, (c)$I_{\text{in}}=230$ and (d)$I_{\text{in}}=250$, indicated by the
white dashed lines in left panel. The parameters are $\protect\kappa %
_{a}=0.1 $, $\protect\kappa _{b}=0.0002$, $\protect\kappa _{\text{ex}}=0.1$,
$\protect\delta =0$, $g_{1}=0.001$, $g_{2}=0.02$, $\protect\omega _{m}=1$
and $\protect\gamma _{m}=0.001$.}}
\label{figure3}
\end{figure}

In figures \ref{figure3}(a)-(d), the transmission $T$ is plotted versus the
detuning $\Delta $ with different input light intensities. It is seen that,
for a small driving intensity, there is a symmetric narrow transparency
window in the middle of transmission spectrum (figure \ref{figure3}(a)),
which is very similar to the typical CRIT effect except for a slight shift
of the center of the transparency window. The result can be explained as
follows, for a tiny input intensity, the value of the mechanical resonator
position can be approximately rewritten as
\begin{equation}
x_{\text{s0}}=g_{1}\kappa _{\text{ex}}I_{\text{in}}/\left( \omega
_{m}g_{2}^{2}\right)
\end{equation}%
which is independent of $\Delta $. Therefore the mechanical motion merely
shifts the center without changing the shape of the transparency window.
With $I_{\text{in}}$ increasing, the transparency window bends towards the
right side and the transmission curve becomes bistable (figure 3(b)). It is
interesting that, for certain $I_{\text{in}}$, the distorted transparency
window breaks into a mother branch and an isola (figure \ref{figure3}(c)).
With further increasing $I_{\text{in}}$, the transparency window becomes
severely distorted, even intersecting with itself and forming a closed loop
(figure \ref{figure3}(d)). From Eqs. (\ref{b}) and (\ref{xs}) we can understand
this phenomenon by noting that the steady-state position of the mechanical resonator
shifts to right-hand side in a way that depends on the detuning for large $I_{\text{%
in}}$ (shown in the left panel of figure \ref{figure3}). For large enough
input light intensity, the nonlinear shift causes an instability in intracavity
photon number and system variables. These effects are dramatically
different from the original CRIT effect, in which the transparency window is
totally independent of the input light intensity and the shape is always
symmetric.

\textit{Weak force detection.} It is well known that, near transparency
windows of EIT or similar effects like CRIT and OMIT, the dispersion is very
large in the vicinity of the non-absorptive resonance point. Thus, a small
detuning from the resonance will lead to a huge phase shift, which can be
used for sensing small perturbations, for instance, ultra-sensitive
detection of magnetic field \cite{PhysRevLett.69.1360,Lee1998Sensitive}.
Here we utilize the CRIT effect and its dependence on optomechanical
nonlinearity to design a weak force detector.

\begin{figure}[th]
\includegraphics[width=3.7in]{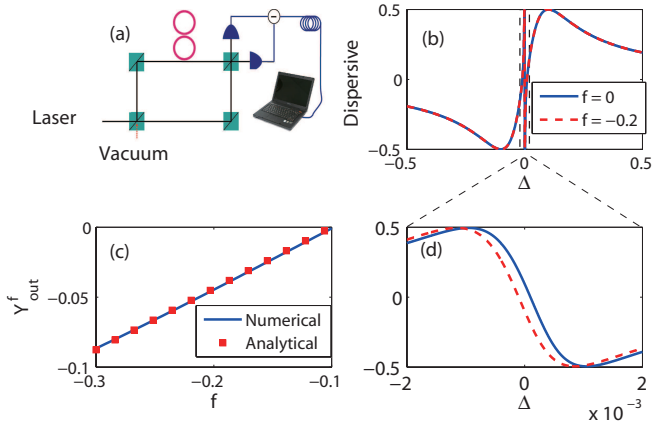}
\caption{{\protect\footnotesize (Color online) (a) The scheme for weak force
detection by measuring phase shifts when a light passing through the hybrid
CRIT-optomechanical system. (b) The dispersion of the system as a function
of detuning $\Delta $ with (red dashed line) and without (blue solid blue
line) external forces. Figure (d) is the zooming in of the labeled region of
figure (b). (c) The phase shift as a function of the external force $f$. The
blue solid line presents the numerical result and the red dashed line is the
approximate analytical result based on Eq. (\protect\ref{Yf}). The
parameters are $\protect\kappa _{a}=0.1$, $\protect\kappa _{b}=0.00002$, $%
\protect\kappa _{\text{ex}}=0.1$, $\protect\delta =0$, $\Delta =0$, $%
g_{1}=0.001$, $g_{2}=0.01$, $I_{in}=0.1$, $\protect\omega _{m}=1$ and $%
\protect\gamma _{m}=0.1$.} }
\label{figure4}
\end{figure}

Assuming that an external force $f$ is applied on the toroidal optomechanical system in our model
and remembering that $\kappa _{\text{ex}}g_{1}x_{s}\ll g_{2}^{2}$, we
evaluate the new steady-state mechanical position involving the external
force $f$ as
\begin{equation}
x_{s}=\omega _{m}^{-1}f+x_{\text{s0}}  \label{xsf}
\end{equation}
The external force shifts the steady-state position of mechanical resonator
from $x_{\text{s0}}$ to $x_{s}$, and thereby shifts the resonance of
CRIT (see the dispersion spectrum in figure \ref{figure4}(b)). With
appropriate parameters chosen, the slope of the dispersion curve in the
vicinity of the resonance is huge. A weak external force can lead to a large
phase shift along with small absorption of the input field, with an
interferometric method, as shown in figure \ref{figure4}(a). This is the key
idea of our proposal for weak force detection.

In figure \ref{figure4}(c) the solid curve represents the
relation between the phase quadrature of the output field $Y_{\text{out}%
}^{f} $ and $f$. One can see that, in the range of the plot, the phase
quadrature linearly depends on $f$, which is ideal for measurements in practice. To analytically verify this
conclusion, we find that, given the steady-state value of the mechanical
resonator position very tiny, the phase quadrature can be approximated as
\begin{equation}
Y_{\text{out}}^{f}=\frac{\sqrt{2I_{\text{in}}}\kappa _{\text{ex}}g_{1}}{%
\omega _{m}g_{2}^{2}}\left( f+\omega _{m}x_{\text{s0}}\right) .  \label{Yf}
\end{equation}%
From figure \ref{figure4}(c), it is clear that the approximate analytical result (%
\ref{Yf}) well agrees with the numerical one.

Now we estimate the sensitivity of our force detection proposal. When the
hybrid CRIT-optomechanics system operates at the resonant point, we
transform the linearized quantum Langevin equations of motion to the
frequency domain (see the supplementary material). According to the
input-output relation, the phase quadrature of the output field in the
homodyne detection can be presented as $Y_{\text{out}}\left( \omega \right)
=Y_{\text{in}}\left( \omega \right) -\sqrt{\kappa _{\text{ex}}}Y_{a}\left(
\omega \right) $. To calculate the sensitivity to the external force $f$, we
define an effective force noise:
\begin{equation}
F\left( \omega \right) =\left. \frac{Y_{\text{out}}\left( \omega \right) }{%
\partial Y_{\text{out}}\left( \omega \right) /\partial f}\right\vert _{f=0}.
\end{equation}%
The total power spectral density of the effective force in the homodyne
measurement of the phase quadrature is%
\begin{eqnarray}
S_{\mathrm{FF}} &=&\int d\omega ^{\prime }\left\langle F\left( \omega
\right) F\left( \omega ^{\prime }\right) \right\rangle  \\
&=&S_{\mathrm{FF}}^{\mathrm{th}}+S_{\mathrm{FF}}^{\mathrm{shot}}  \notag
\end{eqnarray}%
where the thermal noise spectral density is $S_{\mathrm{FF}}^{\mathrm{th}%
}=2m\gamma _{m}K_{B}T_{R}$ and the dimensional optical shot noise spectral density, for
a Dc ($\omega =0$) force, is given by
\begin{equation}
S_{\mathrm{FF}}^{\mathrm{shot}}\left( 0\right) \simeq \frac{\hbar m\omega
_{m}^{2}}{4}\left[ \frac{1}{2}\frac{g_{2}^{6}}{\left( \kappa
g_{1}x_{s}\right) ^{3}}-\frac{g_{2}^{2}}{\left( \kappa g_{1}x_{s}\right) }+%
\frac{9}{2}\frac{\left( \kappa g_{1}x_{s}\right) }{g_{2}^{2}}\right]  \label{S}
\end{equation}%
Here we have assumed $\kappa _{a},\kappa _{b}\ll \kappa _{\text{ex}%
},g_{2}$. The power spectral density or the
square of the force sensitivity is minimized at $g_{2}^{2}\simeq 1.45\kappa g_{1}x_{s}$:
\begin{eqnarray}
\mathrm{min}[S_{\mathrm{FF}}^{\mathrm{shot}}]\simeq 0.8\hbar m\omega _{m}^{2}
\end{eqnarray}
This is below the standard quantum limit. Reasonable system parameters are assumed as: $\omega _{m}\sim 20$ MHz, $m\sim 9$ pg, $\gamma _{m}\sim 40$ Hz, $g_{1}\sim 3$ MHz$\cdot $nm$^{-1}$, $%
g_{2}\sim 4.6$ MHz, $\kappa _{1}\sim 1$ MHz, $\kappa _{2}\sim 0.01$ MHz, $%
\kappa _{\text{ex}}\sim 200$ MHz, the optimal input power $I_{\text{in}}\sim
10$ $\mu $W. Using these parameters, the estimated optimal force sensitivity is $\sqrt{S_{\mathrm{FF}}^{\mathrm{shot}}}%
\sim 17$ aN$\cdot $Hz$^{-1/2}$ within 1$s$ averaging time. This result is comparable to the experimental result reported in \cite{Gavartin2012A} but with shorter averaging time and without using feedback control.

\textit{Conclusions.} We have investigated a CRIT system with one of the
optical ring resonators replaced by a toroidal optomechanical resonator. The
bistability of the light transmission and the equilibrium position of
mechanical resonator were studied in the semiclassical limit. Interestingly,
there are isolas and self-sustained oscillation dynamics appearing in the
system. The nonlinearity induced by the optomechanical resonator on the CRIT
effect was also studied. The result shows that the transparency window is
dramatically affected by the optomechanical coupling. Finally, we suggested
a weak force detection scheme based on the described system and found that
the optimal force sensitivity of $17$ aN$\cdot $Hz$^{-1/2}$ with available experimental
conditions could be achieved.

Z.D. and B.F. acknowledge support from National Natural Science Foundation of China through Grant No. 11364021, No. 11504145 and No. 61368001, and Natural Science Foundation of Jiangxi Province through Grant No. 20122BAB212005.

\section{Appendix}
In this appendix we give the detailed calculation for evaluating the
force detection sensitivity.

The linearized equations of motion for the system variables in the frequency
domain are given by
\begin{eqnarray}
x &=&\chi _{m}\left[ \xi _{m}+f+\sqrt{2}\hbar g_{1}b_{s}X_{b}\right] \\
\left( \kappa -i\omega \right) X_{a} &=&g_{2}Y_{b}+\sqrt{\kappa _{\text{ex}}}%
X_{\text{d}}+\sqrt{\kappa _{a}}X_{\text{ac}} \\
\left( \kappa -i\omega \right) Y_{a} &=&-g_{2}X_{b}+\sqrt{\kappa _{\text{ex}}%
}Y_{\text{d}}+\sqrt{\kappa _{a}}Y_{\text{ac}} \\
\left(\frac{ \kappa _{b}}{2}-i\omega \right) X_{b} &=&-g_{1}x_{s}Y_{b}+g_{2}Y_{a}+%
\sqrt{\kappa _{b}}X_{\text{bc}} \\
\left( \frac{\kappa _{b}}{2}-i\omega \right) Y_{b} &=&g_{1}x_{s}X_{b}+\sqrt{2}%
g_{1}b_{s}x-g_{2}X_{a}\\\nonumber
&+&\sqrt{\kappa _{b}}Y_{\text{bc}}
\end{eqnarray}
where $\chi _{m}^{-1}=m\left( \omega _{m}^{2}-\omega ^{2}-i\gamma _{m}\omega
\right) $ is the mechanical susceptibility. For simplicity we choose the
initial condition of the system such that the phase of the optical field b
is zero, i.e., $b_{s}$ is real.

According to the input-output relation, the phase quadrature of the output
field is $Y_{\text{out}}=Y_{\text{d}}-\sqrt{\kappa _{\text{ex}}}Y_{a}$.
After some calculation we arrive at
\begin{eqnarray}
Y_{\mathrm{out}}&=&\frac{\chi _{\mathrm{b}}g_{2}\sqrt{\kappa _{\mathrm{ex}}}}{
\left( \kappa -i\omega \right) }( \chi _{\mathrm{F}}(\xi _{m}+f)+\chi _{
\mathrm{Xd}}X_{\mathrm{d}}\\\nonumber
&+&\chi _{\mathrm{Xac}}X_{\mathrm{ac}}
+\chi _{
\mathrm{Ybc}}Y_{\mathrm{bc}}+\chi _{\mathrm{Yd}}Y_{\mathrm{d}}+\chi _{
\mathrm{Yac}}Y_{\mathrm{ac}}+\chi _{\mathrm{Xbc}}X_{\mathrm{bc}})
\end{eqnarray}
where
\begin{eqnarray}
\chi _{F} &=&-\sqrt{2}\chi _{m}g_{1}x_{s}g_{1}b_{s}\left( \kappa -i\omega
\right) ^{2} \\
\chi _{\mathrm{Xd}} &=&g_{1}g_{2}x_{s}\sqrt{\kappa _{\text{ex}}}\left(
\kappa -i\omega \right)  \\
\chi _{\mathrm{Yd}} &=&\sqrt{\kappa _{\text{ex}}}\left( Ag_{2}+\frac{\left(
\kappa -\kappa _{\text{ex}}-i\omega \right) }{\chi _{b}g_{2}\kappa _{\text{ex%
}}}\right)  \\
\chi _{\mathrm{Xac}} &=&g_{1}g_{2}x_{s}\sqrt{\kappa _{a}}\left( \kappa
-i\omega \right)  \\
\chi _{\mathrm{Yac}} &=&\sqrt{\kappa _{a}}\left( Ag_{2}-\frac{1}{g_{2}\chi
_{b}}\right)  \\
\chi _{\mathrm{Xbc}} &=&A\sqrt{\kappa _{b}}\left( \kappa -i\omega \right)  \\
\chi _{\mathrm{Ybc}} &=&-g_{1}x_{s}\sqrt{\kappa _{b}}\left( \kappa -i\omega
\right) ^{2}
\end{eqnarray}%
and%
\begin{eqnarray}
A &=&\left( \kappa _{b}/2-i\omega \right) \left( \kappa -i\omega \right)
+g_{2}^{2} \\
\chi _{b}^{-1} &=&A^{2}+\left( g_{1}x_{s}\right) ^{2}\left( \kappa -i\omega
\right) ^{2}\left( 1+2m\omega _{m}^{2}\chi _{m}\right)
\end{eqnarray}%
where $\xi _{m}$ obeys the correlation $\left\langle {\xi _{m}(t)\xi
_{m}(t^{\prime })}\right\rangle =2m\gamma _{m}K_{B}T_{R}\delta (t-t^{\prime
})$ with Boltzmann constant $K_{B}$ and the environment temperature $T_{R}$.
To calculate the sensitivity of force detection, an effective force noise is
defined as
\begin{eqnarray}
F\left( \omega \right)  &=&\left. \frac{Y_{\text{out}}\left( \omega \right)
}{\partial Y_{\text{out}}\left( \omega \right) /\partial f}\right\vert _{f=0}
\\\nonumber
&=&\xi _{m}+( \chi _{Xd}X_{\text{d}}+\chi _{Xac}X_{\text{ac}}+\chi
_{Ybc}Y_{\text{bc}}+\chi _{Yd}Y_{\text{d}}\\\nonumber
&+&\chi _{Yac}Y_{\text{ac}}+\chi
_{Xbc}X_{\text{bc}}) /\chi _{F}
\end{eqnarray}%
The total power spectral density(PSD) is%
\begin{eqnarray}
S_{\mathrm{FF}}\left( \omega \right)  &=&\int d\omega ^{\prime }\left\langle
F\left( \omega \right) F\left( \omega ^{\prime }\right) \right\rangle  \\\nonumber
&=&S_{\mathrm{FF}}^{\mathrm{th}}\left( \omega \right) +S_{\mathrm{FF}}^{%
\mathrm{shot}}\left( \omega \right)
\end{eqnarray}%
where the thermal noise PSD is%
\begin{eqnarray}
S_{\mathrm{FF}}^{\mathrm{th}}\left( \omega \right)  &=&\int d\omega ^{\prime
}\left\langle \xi \left( \omega \right) \xi \left( \omega ^{\prime }\right)
\right\rangle  \\\nonumber
&=&2m\gamma _{m}K_{B}T
\end{eqnarray}%
and the shot noise PSD is%
\begin{eqnarray}
S_{FF}^{\mathrm{shot}}\left( \omega \right) &=&\frac{1}{2}\left\vert \frac{%
\chi _{Xd}-i\chi _{Yd}}{\chi _{F}}\right\vert ^{2}+\frac{1}{2}\left\vert
\frac{\chi _{Xac}-i\chi _{Yac}}{\chi _{F}}\right\vert ^{2}\\\nonumber
&+&\frac{1}{2}%
\left\vert \frac{\chi _{Xbc}-i\chi _{Ybc}}{\chi _{F}}\right\vert ^{2}
\end{eqnarray}

For a Dc $\left( \omega =0\right) $ force, we have%
\begin{eqnarray}
\left\vert \frac{\chi _{\mathrm{Xd}}-i\chi _{Yd}}{\chi _{F}}\right\vert ^{2}
&=&\frac{\hbar m\omega _{m}^{2}}{2}\frac{\kappa _{\text{ex}}}{\kappa }\frac{%
\left( \kappa g_{1}x_{s}\right) ^{2}g_{2}^{2}+\frac{\left( g_{2}^{4}-3\left(
\kappa g_{1}x_{s}\right) ^{2}\right) ^{2}}{4g_{2}^{2}}}{\left( \kappa
g_{1}x_{s}\right) ^{3}} \\
\left\vert \frac{\chi _{\mathrm{Xac}}-i\chi _{\mathrm{Yac}}}{\chi _{F}}%
\right\vert ^{2} &=&\frac{\hbar m\omega _{m}^{2}}{2}\frac{\kappa _{\text{a}}%
}{\kappa }\frac{\left( \kappa g_{1}x_{s}\right) ^{2}g_{2}^{2}+\frac{9\left(
\kappa g_{1}x_{s}\right) ^{4}}{g_{2}^{2}}}{\left( \kappa g_{1}x_{s}\right)
^{3}} \\
\left\vert \frac{\chi _{\mathrm{Xbc}}-i\chi _{\mathrm{Ybc}}}{\chi _{F}}%
\right\vert ^{2} &=&\frac{\hbar m\omega _{m}^{2}}{2}\frac{\kappa _{\text{b}}%
}{\kappa }\kappa ^{2}\frac{g_{2}^{4}+\left( g_{1}x_{s}\kappa \right) ^{2}}{%
\left( \kappa g_{1}x_{s}\right) ^{3}}.
\end{eqnarray}%
here we have assumed $\kappa _{a},\kappa _{b}\ll g_{2},\kappa _{\text{ex}}$. Then, the noise PSD becomes
\begin{equation}
S_{\mathrm{FF}}^{\mathrm{shot}}\left( 0\right) \simeq \frac{\hbar m\omega
_{m}^{2}}{4}\left[ \frac{1}{2}\frac{g_{2}^{6}}{\left( \kappa
g_{1}x_{s}\right) ^{3}}-\frac{g_{2}^{2}}{\left( \kappa g_{1}x_{s}\right) }+%
\frac{9}{2}\frac{\left( \kappa g_{1}x_{s}\right) }{g_{2}^{2}}\right] ,
\label{S3}
\end{equation}%
When $g_{2}^{2}\simeq 1.45\kappa g_{1}x_{s}$ is satisfied, the power spectra
density can be minimized
\begin{equation}
S_{\mathrm{FF}}^{\mathrm{shot}}\left( 0\right) \simeq 0.8\hbar m\omega
_{m}^{2}
\end{equation}%
This is the square of the force detection sensitivity

\bibliographystyle{apsrev4-1}

%\bibliography{CRIT_optomech}
\end{document}